\documentclass[a4paper,10pt, twocolumn]{article} 

\usepackage[a4paper, margin=0.8in]{geometry} 
\usepackage{graphicx} 
\usepackage{amsmath, amssymb} 
\usepackage{hyperref} 
\usepackage{booktabs} 
\usepackage{caption} 
\usepackage{setspace} 
\usepackage{titlesec} 
\usepackage{listings} 
\usepackage{xcolor} 
\usepackage{titling}

\titleformat{\section}{\large\bfseries}{\thesection}{1em}{}
\titleformat{\subsection}{\normalsize\bfseries}{\thesubsection}{1em}{}

\definecolor{dkgreen}{rgb}{0,0.6,0}
\definecolor{gray}{rgb}{0.5,0.5,0.5}
\definecolor{mauve}{rgb}{0.58,0,0.82}
\definecolor{eclipseStrings}{RGB}{42,0.0,255}
\definecolor{eclipseKeywords}{RGB}{127,0,85}

\colorlet{numb}{magenta!60!black}

\lstdefinelanguage{json}{
	basicstyle=\normalfont\ttfamily,
	commentstyle=\color{eclipseStrings}, 
	stringstyle=\color{eclipseKeywords}, 
	numbers=left,
	numberstyle=\scriptsize,
	stepnumber=1,
	numbersep=8pt,
	showstringspaces=false,
	breaklines=true,
	frame=lines,
	string=[s]{"}{"},
	comment=[l]{:\ "},
	morecomment=[l]{:"},
	literate=
	*{0}{{{\color{numb}0}}}{1}
	{1}{{{\color{numb}1}}}{1}
	{2}{{{\color{numb}2}}}{1}
	{3}{{{\color{numb}3}}}{1}
	{4}{{{\color{numb}4}}}{1}
	{5}{{{\color{numb}5}}}{1}
	{6}{{{\color{numb}6}}}{1}
	{7}{{{\color{numb}7}}}{1}
	{8}{{{\color{numb}8}}}{1}
	{9}{{{\color{numb}9}}}{1}
}

\title{\textbf{Sustainable Open-Data Management for Field Research: A Cloud-Based Approach in the Underlandscape Project} }
\author{
	Ciuffoletti, Augusto \\
	University of Pisa \\ \texttt{augusto.ciuffoletti@unipi.it}
	\and
	Chiti, Letizia \\ 
	University of Parma \\
	\texttt{letizia.chiti@unipr.it}
}
\date{\today}

\begin{document}
	
	\maketitle 
	
	\let\thefootnote\relax\footnotetext{This research was conducted as part of the \href{https://sites.google.com/view/prin-underlandscape}{Underlandscape project}, a PRIN research initiative funded by the Italian Ministry of University and Research (MUR).}
	
	\begin{abstract}
		\noindent 
		Field-based research projects require a robust suite of ICT services to support data acquisition, documentation, storage, and dissemination. A key challenge lies in ensuring the sustainability of data management—not only during the project's funded period but also beyond its conclusion, when maintenance and support often depend on voluntary efforts.
		
		In the Underlandscape project, we tackled this challenge by extensively leveraging public cloud services while minimizing reliance on complex custom infrastructure.
		
		This paper provides a comprehensive overview of the project's final infrastructure, detailing the adopted data formats, the cloud-based solutions enabling data management, and the custom applications developed for system integration. It complements the poster \cite{ciu25a}, which was on display during the \href{https://sites.google.com/view/prin-underlandscape/workshop/workshop-the-shadows-in-the-cave}{Underlandscape Workshop} held in Pisa in January 2025.
	\end{abstract}
	
	\section{Introduction}
	
	Field-based scientific research projects generate extensive datasets that demand efficient management at every stage of the project lifecycle. The infrastructure supporting such projects must enable seamless data acquisition, documentation, processing, storage, and dissemination while ensuring long-term accessibility. The challenge lies not only in designing a system that meets the needs of researchers and stakeholders during the project's active phase but also in sustaining data availability after formal funding has ended.
	
	Traditionally, research projects rely on dedicated servers or institutional repositories for data management. However, these solutions often face sustainability issues, as ongoing maintenance requires dedicated personnel and financial resources. Without a structured approach, valuable data risk becoming inaccessible once a project concludes, diminishing the long-term impact of the research.
	
	The Underlandscape project presents a model for addressing these challenges through a cloud-based infrastructure. By leveraging widely available public cloud services, the project minimizes the need for custom-built software while ensuring scalability, reliability, and ease of maintenance. This approach reduces operational costs and enhances data longevity, making it possible to keep resources accessible even after the project's funding period.
	
	This paper outlines the technical decisions made in the Underlandscape project to establish a sustainable data management framework. We begin by defining the data formats that facilitate interoperability and long-term usability. Next, we describe the cloud services employed to support secure storage and easy retrieval of information. Finally, we discuss the development of lightweight applications that integrate different components, ensuring smooth data flow without introducing unnecessary complexity.
	
	By sharing our experience, we aim to provide a reference model for other research projects facing similar sustainability concerns. Our approach demonstrates that with careful planning and strategic use of cloud technologies, research data can remain accessible and useful beyond the lifespan of a single project.
	
	\section{Methods}
	
	The Underlandscape project is a multidisciplinary and diachronic study of Tuscan mountain landscapes, initiated in 2022 by the CNR and the University of Pisa. It aims to establish an analysis protocol for the study and conservation of rock structures—from prehistoric to contemporary times—using non-invasive scientific techniques. Given the cultural and environmental significance of these structures, as well as their historical and archaeological potential, the project also seeks to develop sustainable strategies for their valorization.
	
	From the project's objectives, we identify four key tasks necessary for implementing the supporting ICT infrastructure:
	
	\begin{enumerate}
		\item The design of an efficient data format to facilitate seamless data flow within the infrastructure.
		\item Establishing a long-term storage solution to ensure content accessibility beyond the project's duration.
		\item Developing a user-friendly interface for accessing the content, supporting multiple approaches to these tasks.
		\item Providing tools for content creation and editing, accommodating diverse modes of engagement.
	\end{enumerate}
	
	Given the project's inherently multidisciplinary nature, the implementation of these components must account for the diverse expertise within the team, which includes specialists in physics, archaeology, botany, and social sciences. While team members possess general computer literacy, they may lack specialized technical knowledge in ICT systems. Therefore, the tools must balance functionality with accessibility, abstracting technical complexities and offering intuitive interfaces that facilitate content management.
    
    \subsection{The data format}
    
	The content produced by the Underlandscape project is predominantly geolocalized. The primary dataset consists of survey documentation, complemented by intermediate waypoints and points of interest relevant to tourism. Given this strong spatial component, we required a data format specifically suited for geolocation applications.
	
	Rather than defining a new format, which would have introduced unnecessary complexity, we opted for an established and widely supported standard: GeoJSON. GeoJSON is a structured extension of JSON, designed explicitly for representing geographical data. Its schema is inherently flexible, allowing for the inclusion of custom attributes without compromising compatibility with existing tools and libraries.
	
	The project's data repository is structured as a collection of GeoJSON files. Initially, we considered storing the data in a non-relational database, given the hierarchical nature of geospatial information. However, we ultimately determined that such an approach would introduce unnecessary overhead and increase maintenance complexity, particularly for non-specialist users. By organizing the data as standalone GeoJSON files, we ensure ease of access, portability, and long-term maintainability without the need for specialized database management skills.
	
	Each file—referred to as an \textit{Underlandscape dataset}—contains a single GeoJSON \texttt{FeatureCollection}, a structured JSON object. The \texttt{properties} field, which is left unspecified by the GeoJSON standard, provides metadata about the dataset and links it to the broader infrastructure. It contains the following key properties:

	\begin{itemize}
		\item \textbf{Nome}—a unique string identifier for the dataset.
		\item \textbf{Descrizione}—a detailed textual description of the dataset’s contents and purpose.
		\item \textbf{umapKey}—a URL pointing to a corresponding uMap map for visualization.
		\item \textbf{WebPageURL}—a link to a related webpage on the project’s website for additional information.
	\end{itemize}

	By definition, a \texttt{FeatureCollection} contains an array of GeoJSON features. Each feature includes structured attributes such as \texttt{type} and \texttt{geometry}, along with a \texttt{properties} object, which is left unspecified in the GeoJSON standard. To ensure consistency across the infrastructure, we define a structured format for feature properties, specifying a set of keys that should be included.

	The dataset format defines six distinct types of features. The type of a given feature is determined by the value associated with the \texttt{ulsp\_type} property within the \texttt{properties} object:

	\begin{itemize}
		\item \texttt{Sito}—a \texttt{Point} feature representing a site of archaeological interest. This is the most detailed feature type, containing additional archaeology-related data. It serves as a summary of survey findings.
		\item \texttt{POI}—a \texttt{Point} feature representing a Point of Interest (POI). It includes a textual description, an image, and categorization tags. Like a \texttt{Sito}, it is an outcome of surveys but with less detailed information.
		\item \texttt{QRtag}—a \texttt{Point} feature representing a QR code-based sign. It contains a textual description and is used for on-site information dissemination, as described in \cite{der23a}.
		\item \texttt{Risorsa}—a \texttt{Point} feature representing a resource relevant to the touristic valorization of the area, such as a restaurant or museum.
		\item \texttt{Percorso}—a \texttt{MultiLineString} feature representing the path followed during a survey.
		\item \texttt{Itinerario}—a \texttt{MultiLineString} feature defining a recommended route for exploring the area, intended for tourism and valorization purposes.
	\end{itemize}

	The dataset property schema is comprehensively documented in a structured JSON object stored in the \texttt{formats.js} file. This file serves both as a reference for documentation and as a resource for custom applications. It is available in the project's GitHub \href{https://github.com/prin-underlandscape/OfflineForm}{repository}.
	
	\subsection{The storage}
	
	All datasets generated within the Underlandscape project are stored in a single Git repository hosted on GitHub (\href{https://github.com/prin-underlandscape/Master}{here}).
	
	This design choice represents a deliberate tradeoff between using a traditional database—whether relational or non-relational—and adopting a simpler, file-based storage system. While database solutions provide powerful query capabilities and sophisticated data management features, they require a dedicated and robust interface for user interaction. Moreover, operating a database in offline conditions presents significant challenges, particularly in a fieldwork setting where internet access may be intermittent or unavailable.
	
	Conversely, our application scenario demands only basic data management operations, such as editing, adding, and updating content. Additionally, our team consists of experts from diverse disciplines, including archaeology, physics, botany, and social sciences, many of whom are not specialists in database management. Therefore, we sought a solution that would be intuitive and accessible to all contributors. Managing the contents of a structured directory aligns well with the team's skill set, and working with a Git repository remains relatively straightforward as long as it follows a basic workflow: modifying files locally, committing changes (typically replacing or updating files), pushing updates to the remote repository, and occasionally retrieving previous file versions to correct errors. This workflow enables seamless collaboration without the complexities of database administration.
	
	To facilitate contributions from multiple team members, Git provides the \textit{pull request} mechanism. A non-registered collaborator can download a copy of the repository (hereafter referred to as the \textit{master}), modify it locally, and then submit a request for the repository manager to integrate the changes. The manager reviews the proposed modifications and merges them into the master repository if they are valid. These tasks can be performed either through a user-friendly graphical interface, such as the \href{https://docs.github.com/en/desktop}{GitHub Desktop} application, or via command-line tools.
	
	In practice, the Underlandscape project employs an even simpler workflow for managing its more than 40 datasets. Instead of requiring contributors to interact with Git directly, they submit new versions of their datasets to a designated database manager. The manager verifies the validity of the data and executes the necessary commit-push sequence from the command line. This approach reduces the technical burden on contributors while ensuring that all datasets remain up to date and accessible. Contributors can always find the latest version of their datasets online, along with a complete history of updates for reference.
	
	Despite its effectiveness, the storage solution described above does not provide a meaningful way to present and communicate the dataset contents. This limitation is particularly evident in a multidisciplinary context, where accessibility and usability are crucial. To address this, we implemented additional solutions to automatically generate and deliver meaningful views of the data directly from the Master repository, which will be illustrated in the following sections.
	
	\subsection{Dataset life-cycle}
	
	The dataset management follows a three steps life-cycle:
	\begin{itemize}
		\item \textbf{creation} -- which may happen either on-site, typically during a survey, or desk-based
		\item \textbf{editing} -- multistep refinement of the content of the dataset
		\item \textbf{delivery} -- render the content of the dataset so that others can take advantage of it
	\end{itemize}
		
	The following sections explore the various modalities to perform such activities we have experimented and implemented in the Underlandscape project.
	
	\subsubsection*{On-Site Dataset Creation}
	
	A key characteristic of on-site dataset creation is the availability of geographical coordinates via GPS, while internet connectivity is often unavailable. We have tested two approaches to address this challenge.  
	
	The first approach leverages \href{https://www.gaiagps.com}{GaiaGPS}, a smartphone app designed for outdoor navigation. Additionally, we evaluated \href{https://qfield.org/}{QField} and \href{https://www.geopaparazzi.org/smash/}{SMASH}, both offering field data collection and geographic documentation features. QField, while a powerful tool for integrating with QGIS, proved impractical for our specific scenario due to its heavy app size, the complexity of loaded layers, and its rigid workflow, which requires prior QGIS project processing. SMASH, though more lightweight and versatile, presented challenges in the data retrieval and processing phase, ultimately leading us to discard it. However, its approach to structuring data collection inspired the adoption of a JSON-based form description. This comparative analysis guided our selection of the most effective tool for our research methodology.
	
	GaiaGPS allows users to record their movements, mark relevant locations with associated textual notes or images, and store this information locally. Once an internet connection becomes available, the recorded data can be uploaded to the user's cloud space and subsequently downloaded in GeoJSON format. While Gaia GPS offers many additional features, those described above are the most relevant for surveys.  
	
	However, the exported GeoJSON files do not conform to the dataset format used in our project. To address this, we developed a JavaScript script, \href{https://github.com/prin-underlandscape/OfflineForm/blob/main/gaiaSplit.js}{gaiaSplit.js}, which restructures the Gaia GPS output into a standardized dataset while preserving recorded positions, textual notes, and images.  
	
	While this method is sufficient in most cases, certain sites of particular interest require a more structured and exhaustive data collection process. To facilitate this, we developed a dedicated mobile application, \textit{ULSP\_App} (\href{https://github.com/prin-underlandscape/ULSP_App/tree/main}{source code and .apk available here}). This app allows users to complete a detailed form that automatically includes GPS coordinates. The default form, designed specifically for archaeological survey, contains over 20 fields. Moreover, since the form structure is defined in the \texttt{formats.js} file, the team's ICT specialist can easily modify it or add new formats tailored to other research needs.  
	
	The \textit{ULSP\_App} operates entirely offline. To enable data transfer between devices, such as laptops or tablets, it uses a QR code-based exchange system. The app generates QR codes containing dataset information, which can be displayed on-screen for scanning, while input data can be captured using the smartphone’s camera.  
	
	\subsubsection{Desk-Based Dataset Creation}
	
	Since the Underlandscape dataset format extends GeoJSON, any valid GeoJSON file can serve as the foundation for a dataset. However, only features conforming to the dataset’s specifications will be fully processed, while any additional properties outside the defined schema will be ignored. As a result, any tool capable of exporting data in GeoJSON format can be used to initiate the dataset lifecycle. Examples include online services such as \href{https://umap.openstreetmap.fr/}{uMap} and desktop applications like \href{https://qgis.org/}{QGIS}.  
	
	Maps available in alternative formats, such as KML or GPX, can be converted to GeoJSON using either online conversion tools or the command-line utility \textit{ogr2ogr}.  
	
	Additionally, datasets can be generated from structured tabular data. Users can create input for the dataset editing tool using an appropriately formatted CSV file, which can be easily produced with any spreadsheet software. The predefined models for each feature type included in the dataset are available \href{https://github.com/prin-underlandscape/OfflineForm/blob/main/formats.xlsx}{here}.
	
	\subsection{Dataset Editing}
	
	The dataset editing process is designed to require only basic ICT skills and to function offline. To meet these requirements, we developed \textit{Off}, a dedicated JavaScript-based editor, available \href{https://github.com/prin-underlandscape/OfflineForm}{here}.  
	
	Once the repository is cloned, the editor becomes accessible through a web browser, making it highly portable across various personal computing devices. If an internet connection is available, the editor can also be accessed online \href{https://sites.google.com/view/prin-underlandscape/dataset/strumenti-offline}{here}, hosted by the \textit{GitHub Pages} service.  
	
	The tool allows users to process multiple files within the same session, facilitating the merging of multiple datasets into a single one. It supports input from various sources, including GeoJSON files, QR codes generated by the \textit{ULSP\_app}, and appropriately structured CSV files.  
	
\begin{figure}
	\centering
	\includegraphics[width=0.7\linewidth]{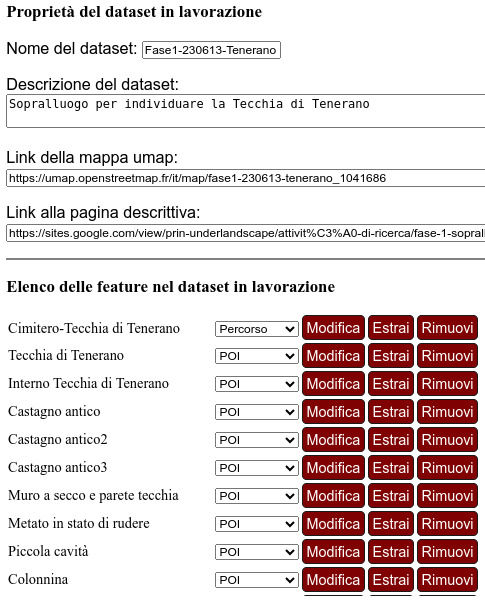}
	\caption[Feature Management in the Off Editor]{The \textit{Off} editor displays a list of dataset features, allowing users to individually remove, edit, or extract them. Feature types can also be modified.}
	\label{fig:off}
	\end{figure}

	The editor displays a list of dataset features (see Figure \ref{fig:off}), allowing users to selectively remove individual elements or retain only specific ones, effectively filtering the dataset. These capabilities offer the flexibility needed to assemble a new dataset from multiple compatible sources.
	
	In addition to these filtering options, users can edit the dataset’s textual description and manage cross-references to the corresponding uMap map and related content on the project website.
	
	The editor also enables users to modify feature types, for instance, transforming a generic \textit{POI} into a data-rich \textit{Sito} to allow for more detailed documentation.
	
\begin{figure}
		\centering
		\includegraphics[width=0.9\linewidth]{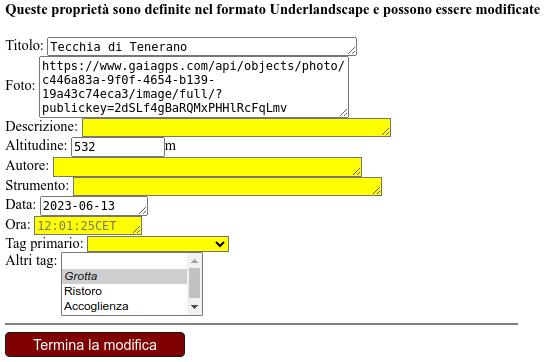}
		\caption[Properties Editing Interface]{The Off editor’s interface for modifying the \textit{Properties} object of a \textit{POI} feature. Users can edit values as free text or select predefined options for certain fields.}
		\label{fig:off-edit}
	\end{figure}

	The editor includes a form-based interface to edit the \textit{Properties} of each feature (see Figure \ref{fig:off-edit}). It displays both recognized properties (in editable text fields) and unrecognized ones (as a separate list). The tool does not automatically discard extraneous properties; instead, it allows users to manually transfer relevant information from unrecognized properties to valid ones before removing the unnecessary data.  
	
	At the end of the editing session, the user saves the updated dataset locally and submits it to the storage manager, who integrates the changes into the master repository by executing a \textit{commit}. Once an internet connection is available, the modifications are \textit{pushed} to the \textit{Master} GitHub repository.  
	
	All steps, except the final push to the remote repository, can be performed entirely offline.  
	
	\subsection{Delivering the Dataset Content}
	
	A dataset provides a structured representation of all available data on a given topic. However, in its raw format, it is not immediately suitable for direct interpretation or user interaction.
	
	To facilitate the visualization and use of datasets, we developed three approaches to render their content effectively.
	
	\subsubsection*{Delivery as a uMap Map}
	
	\href{https://umap.openstreetmap.fr/it/}{uMap} is an online tool that allows users to create custom maps based on \href{http://osm.org/}{OpenStreetMap} layers. It serves as an ideal platform for geographically displaying dataset features, providing an intuitive interface for interacting with spatial information. While uMap offers advanced functionalities for map customization, managing multiple maps in a consistent manner—especially when dealing with rich popups and numerous markers—requires a dedicated automation process.
	
	To address this need, we developed a Python application using the \href{https://www.selenium.dev/}{Selenium} library to automate the generation of maps directly from datasets. The source code for this tool is available \href{umap-sync.py}{here}.
	
	Key features of the generated maps include:
	\begin{itemize}
		\item Representation of \textit{Point} features as markers and \textit{MultiLineString} features as lines.
		\item Automatic organization of markers into layers based on their type.
		\item Popups displaying relevant information extracted from the \textit{Properties} section of each feature.
		\item Custom marker icons assigned based on feature tags when available.
		\item Embedded images and cross-references within popups, linking to the corresponding dataset repository and the relevant page on the project website.
		\item Built-in search capabilities provided by the uMap interface for easier navigation.
	\end{itemize}
	
	To ensure that the maps remain up to date, the database manager routinely runs the \textit{umap-sync.py} script after updating the \textit{Master} repository. This process ensures that all modifications to the dataset are promptly reflected in the visual representation. The automatic synchronization minimizes the risk of outdated information and maintains consistency between the raw datasets and their geographical rendering.
	
\begin{figure}
		\centering
		\includegraphics[width=0.9\linewidth]{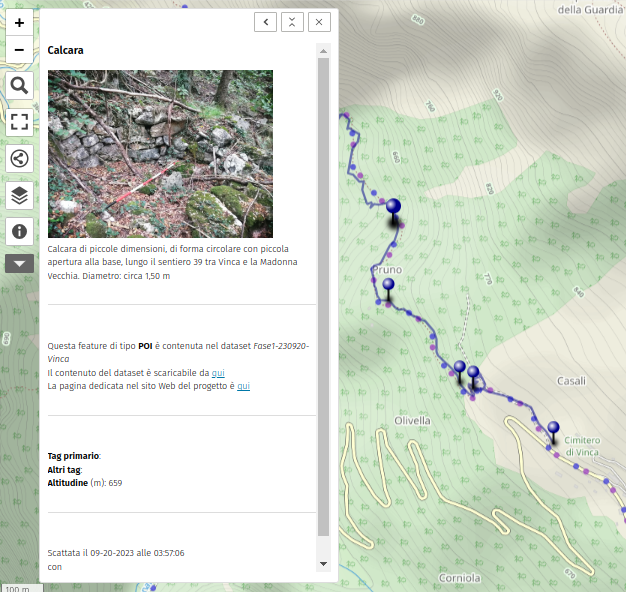}
		\caption[Umap rendering]{Rendering of a dataset on uMap. On the left the popup attached to the highlighed marker. The map is available \href{https://github.com/prin-underlandscape/Fase1-230714-Ghivizzano}{here}}
		\label{fig:umap}
	\end{figure}

	Figure \ref{fig:umap} illustrates an example of a popup attached to a POI feature, which displays key details from the dataset when users interact with the corresponding marker.
	
	In addition to individual dataset-specific maps, the \textit{umap-sync.py} script also generates a global map that consolidates features from all datasets stored in the \textit{Master} repository (\href{https://umap.openstreetmap.fr/it/map/sommario_1044830}{URL}). This summary map provides a comprehensive overview of the project’s data and facilitates comparative analysis across different datasets. Furthermore, uMap’s interface enables users to filter displayed features based on dataset name or tags, enhancing usability and navigation.
	
	To increase accessibility, these dynamically updated maps are embedded directly into relevant pages of the project website, ensuring seamless integration with other digital resources.

	\subsubsection*{Delivery as a Dataset-specific repository}
	
	While geographical visualization offers an intuitive representation of dataset contents, it may present challenges for programmatic processing and structured data management. To address this limitation, we provide an alternative dataset representation in the form of dataset-specific repositories.
	
	Each dataset-specific repository is automatically generated from the dataset using the \href{https://github.com/prin-underlandscape/dataset\_tools/blob/main/repo-sync.py}{repo-sync.py} Python script. This script extracts relevant data from the GeoJSON file and organizes it into a structured repository, which serves as a comprehensive reference for both human users and automated processes.
	
	The generated repository includes, besides the GeoJSON dataset:
	\begin{itemize}
		\item A README file that provides an overview of the dataset contents, including metadata, descriptions, and cross-references to relevant maps on uMap. This ensures that users can quickly grasp the dataset’s scope and significance.
		\item A GPX file that can be loaded into outdoor navigation applications such as \href{https://www.gaiagps.app/}{Gaia GPS} or \href{https://www.locusmap.app/}{Locus Map}, allowing users to access geographic features on mobile devices in offline environments.
		\item A \textit{vignettes} directory containing images linked to dataset features. These images are automatically downloaded from their original online sources and resized to ensure optimal performance and usability. 
		\item A \textit{qrtags} directory, which stores QR tags intended for on-site signage. When applicable, these QR codes facilitate interaction with the dataset by providing direct links to relevant digital content.
	\end{itemize}
	
\begin{figure} 
		\centering 
		\includegraphics[width=0.7\linewidth]{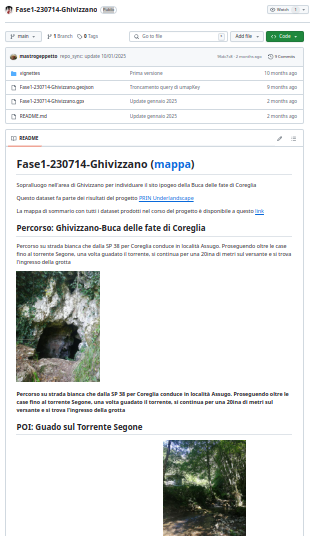}
		\caption[Content of the dataset-specific repository]{A sample of the content of a dataset-specific repository. The complete repository is available \href{https://github.com/prin-underlandscape/Fase1-230714-Ghivizzano}{here}.}
		\label{fig:datasetrepo}
	\end{figure}

	Figure \ref{fig:datasetrepo} presents an example of the structure of a dataset-specific repository, showcasing its key components. The complete repository is accessible \href{https://github.com/prin-underlandscape/Fase1-230714-Ghivizzano}{here}.
	
	This dual approach—combining interactive maps with structured dataset repositories—ensures both usability and long-term sustainability. The repositories facilitate structured data processing, archival storage, and interoperability with external tools, while the maps provide an accessible and engaging way to explore the dataset’s contents.
	
	\subsubsection*{Delivery across an outdoor navigation App}
	
	Research activities generate valuable insights and discoveries that extend beyond the academic community, offering opportunities for broader dissemination. By translating complex findings into accessible narratives, visual representations, or interactive experiences, researchers can engage with non-specialist audiences, fostering a deeper appreciation for the subject matter. This process not only democratizes knowledge but also enriches cultural understanding by bridging the gap between specialized expertise and public awareness. Through exhibitions, digital platforms, educational programs, or storytelling, making research findings available to a wider audience enhances collective cultural appreciation, stimulates curiosity, and promotes interdisciplinary dialogue.
	
	In this spirit—fully aligned with the objectives of the Underlandscape project—we designed a dedicated outdoor navigation app that leverages dataset-driven guidance to help visitors explore a specific region of interest.
	
	The Turista app, described in detail in the companion paper \cite{ciu25d}, is distinguished by its ease of use and its focus on assisting rather than overwhelming the user. It avoids unnecessary options and complexity, prioritizing intuitive interaction.
	
	The app is configured by the user for a specific location through a simple process: downloading a dataset from a URL obtained by scanning a QR code. Once configured, the app presents a single, streamlined interface displaying a map with Points of Interest (POIs) and, where applicable, the trails leading to them. Additionally, it highlights the nearest POI on the map, providing its corresponding image, title, and textual description, ensuring a seamless and informative exploration experience.

	\section{Results}
	
	The complex architecture described in the previous sections has been fully implemented. 
	Most of its components have been extensively tested, although certain aspects could benefit from refinements—particularly the management of the valorization tools, such as the \textit{Turista App}, which has not yet undergone a significant Beta test.
	
	One of the most significant outcomes of this work is the demonstrated flexibility of the architecture. 
	The initial core of the system, developed in 2023, was the \textit{ULSP\_App}, which led to the first definition of the dataset structure, initially featuring only the \textit{Sito} type. 
	Subsequent developments expanded the dataset format and functionality:
	
	\begin{itemize}
		\item The need to process Gaia GPS tracks from previous surveys led to the introduction of two additional feature types: \textit{Percorso} and \textit{POI}.
		
		\item The adoption of GitHub as the primary data storage platform, along with the development of \textit{Off}, the dataset editor, streamlined data management.
		
		\item The integration of QR-tagged signals introduced a new dataset feature type to support field-based interaction.
		
		\item The requirement for a uniform visual style in site maps led to the automation of dataset processing through the \textit{repo\_sync} tool and the creation of dataset-specific GitHub repositories.
		
		\item Enhancements for visitor support resulted in the addition of two more feature types: \textit{Itinerario}, a curated track created \textit{a posteriori}, and \textit{Risorsa}, which provides tagged logistical information.
	\end{itemize}
	
	This progressive development underscores the architecture’s adaptability and sustainability, key objectives of the project.
	
	Further extensions are currently under development. 
	A nearly complete integration of the \href{https://www.zotero.org}{Zotero} bibliography reference manager is in progress, and the inclusion of 3D imaging is under consideration.
	
	As the project nears completion, some components of the architecture may eventually be retired, credentials may be lost, or certain elements may become obsolete. 
	However, the most significant outcomes—particularly the structured datasets and tools—will remain accessible through GitHub, ensuring their long-term availability and usability.
	
	\section{Discussion}
	
	Experimental results are at the core of the scientific method. ICT technology provides ways to store large amounts of high-quality data, but ensuring their long-term availability remains a challenge.
	
	This issue is well explained in \cite{sal21a}, where the authors investigate the case of ecological research. They identify four key challenges: \textit{harmonization}, \textit{biases}, \textit{expertise}, and \textit{communication}. Their approach is significantly broader in scope than ours, but we share and directly address three of these challenges.
	
	Starting with the second challenge, \textit{biases}, this falls outside our current objectives. In their discussion, the authors highlight the risks associated with statistical biases, where certain well-documented areas of study receive disproportionately more data, leading to skewed statistical significance. Our research does not incorporate a statistical perspective, as our dataset is primarily descriptive rather than inferential. Consequently, we do not currently tackle this issue.
	
	The challenge, \textit{harmonization}, refers to the need for a common data representation. This was one of the initial concerns in our system's design. To address it, we introduced the Underlandscape Dataset format, ensuring consistency in data structuring while maintaining adaptability. The format has proven sufficiently flexible to accommodate the evolving nature of our project while remaining grounded in widely accepted standards. Furthermore, this standardized format allows for broader applicability; other projects with similar goals could adopt and extend it in a backward-compatible manner. For example, botanical or environmental datasets could integrate seamlessly into a shared repository, fostering cross-disciplinary collaboration.
	
	The third challenge, \textit{expertise}, as defined by \cite{sal21a}, concerns the development of skills necessary for managing open data. As is often the case in multidisciplinary projects, we face the same issue and address it from two complementary angles. First, we strive to simplify data management as much as possible, minimizing technological barriers for non-expert users. Second, we provide structured documentation and tutorials, accessible through a \href{https://sites.google.com/view/prin-underlandscape/dataset/tutorial}{dedicated section} of the project website, to facilitate effective engagement with the dataset and its tools.
	
	The final challenge, \textit{communication}, pertains to identifying and addressing the needs of various stakeholders involved in open data collection and usage. In \cite{ciu25a}, we examine this issue in depth, defining four key personas and analyzing the potential risks associated with data sharing, particularly in relation to over-tourism in the context of cultural heritage valorization. The main actors we identified are: (1) domain experts conducting fieldwork, such as archaeologists and physicists, (2) technical support personnel, primarily ICT specialists, (3) administrative entities overseeing data governance, and (4) visitors who engage with the data through interpretative and navigational tools. A holistic approach that carefully considers their diverse interests and responsibilities is essential for successful and ethical data sharing.
	
	One aspect that \cite{sal21a} does not explore in depth is the technical sustainability of data integrity. Their assumption of a global, centralized database implies the feasibility of dedicated data management efforts. In contrast, our approach acknowledges the challenges of decentralized data stewardship and the necessity of implementing sustainable practices for long-term preservation and accessibility.
	
	The issue of long-term sustainability in open-access databases is extensively discussed in \cite{cos14a}, where the authors analyze both financial and maintenance-related challenges in the biological domain. Their work provides valuable insights into the economic and logistical aspects of sustaining large-scale data repositories over time.
	
	Regarding financial sustainability, the authors highlight that maintaining a catalog of millions of high-resolution images requires an estimated annual cost of approximately €80,000. While this figure does not directly apply to our case, it underscores a crucial consideration: managing open data is not cost-free. Even in smaller-scale projects like ours, there are implicit costs associated with data administration, infrastructure maintenance, and long-term storage.
	
	In our specific use case, we have adopted a strategy to minimize these costs by reducing administrative overhead and relying on low-maintenance infrastructure. However, we have also ensured that our approach remains scalable. The cloud-based architecture we designed is flexible enough to accommodate larger datasets should the need arise, with cost structures that remain manageable for both institutional and commercial stakeholders. 
	
	For our implementation, we opted for GitHub as our primary data repository, balancing accessibility, cost-effectiveness, and ease of use. Initially, we also considered other cloud-based database solutions, such as MongoDB, which could have provided additional functionalities but at a higher operational cost. Additionally, national research infrastructure initiatives, such as the Italian GARR (\textit{Gruppo per l'Armonizzazione delle Reti della Ricerca}), offer viable alternatives to commercial providers. Unfortunately, administrative constraints prevented us from leveraging these resources, highlighting another non-technical challenge in the sustainable management of open data.
	
	The discussion in \cite{cos14a} reinforces the importance of early-stage planning for financial and logistical sustainability. While our current setup meets our immediate needs, a significant scale-up may require a reassessment of infrastructure choices, given the limits of GitHub repositories. In such a scenario, alternative cloud-based solutions or institutional repositories could be considered to ensure long-term sustainability and scalability. This consideration aligns with the observation that scientific databases tend to emerge and gain stability over time, as their size and reputation reach critical thresholds.
	
	The authors also highlight the challenge of long-term open-data management, emphasizing the risk of declining interest from the initial developers and the necessity of a succession plan spanning decades. While our perspective is less ambitious, we acknowledge the likelihood of extended periods without active management. Nevertheless, the data may continue to be used by individuals who are not directly involved in maintaining it. Once again, the key to resilience lies in relying on stable and persistent infrastructure for core services; in our case, this includes GitHub and uMap.

	\section{Conclusion}
	
	The \href{https://sites.google.com/view/prin-underlandscape/progetto}{Underlandscape project} aims to establish a methodology for conducting, documenting, and enhancing the impact of field-based research. A crucial component of this effort has been the development of an ICT infrastructure that facilitates the collection of activity reports, ensures data accessibility for both experts and non-experts, and guarantees the long-term availability of research outputs beyond the project's conclusion.
	
	As the project progressed—now approaching its completion at the time of writing—new requirements continually emerged, necessitating adaptations and integrations within a flexible framework. Initially, we explored a non-relational database solution, but usability challenges led us to adopt a more pragmatic approach. Our final architecture relies on existing cloud-based services, seamlessly integrated through a set of manageable scripts. This design choice proved to be highly adaptable, allowing the infrastructure to evolve in response to changing demands, which we consider a key success factor.
	
	One of the most significant outcomes is the dataset format we developed. By building upon a widely accepted standard for geographic data, we took a step toward defining a shared method for encoding meaningful geographic features. A broader, coordinated effort in this direction could pave the way for open, semantically enriched, discipline-specific maps—an invaluable resource for a wide range of research fields.

	\bibliographystyle{plainurl} 
	\bibliography{paper} 

\begin{thebibliography}{1}

\bibitem{ciu25d}
A.~Ciuffoletti.
\newblock Designing a geo-tourism app: A principled approach, 2025.
\newblock \href {https://doi.org/http://dx.doi.org/10.13140/RG.2.2.14778.91847}
  {\path{doi:http://dx.doi.org/10.13140/RG.2.2.14778.91847}}.

\bibitem{ciu25a}
Augusto Ciuffoletti.
\newblock Underlandscape project data flow, 2025.
\newblock \href {https://doi.org/10.13140/RG.2.2.24968.94725}
  {\path{doi:10.13140/RG.2.2.24968.94725}}.

\bibitem{cos14a}
Mark~J Costello, Ward Appeltans, Nicolas Bailly, Walter~G Berendsohn, Yde
  de~Jong, Martin Edwards, Rainer Froese, Falk Huettmann, Wouter Los, Jan Mees,
  et~al.
\newblock Strategies for the sustainability of online open-access biodiversity
  databases.
\newblock {\em Biological Conservation}, 173:155--165, 2014.
\newblock \href {https://doi.org/10.1016/j.biocon.2013.07.042}
  {\path{doi:10.1016/j.biocon.2013.07.042}}.

\bibitem{der23a}
Maria~Grazia Deri, Letizia Chiti, and Augusto Ciuffoletti.
\newblock A sustainable approach to tourist signage on heritage trails.
\newblock {\em Sustainability}, 15(23):16251, 2023.
\newblock \href {https://doi.org/10.3390/su152316251}
  {\path{doi:10.3390/su152316251}}.

\bibitem{sal21a}
Roberto Salguero-G{\'o}mez, John Jackson, and Samuel~JL Gascoigne.
\newblock Four key challenges in the open-data revolution.
\newblock {\em Journal of Animal Ecology}, 90(9):2000--2004, 2021.
\newblock \href {https://doi.org/10.1111/1365-2656.13567}
  {\path{doi:10.1111/1365-2656.13567}}.

\end{thebibliography}
	
\end{document}